\documentclass[12pt,preprint,eqsecnum,aps]{revtex4} 
\usepackage{times}
\usepackage{graphicx}
\usepackage[fleqn]{amsmath}
\usepackage{amsfonts}
\usepackage{amssymb}
\usepackage{amsthm}
\usepackage{amsopn}
\usepackage{xspace}
\usepackage{array}
\usepackage{epsfig}
\begin{document}
\title{Quantum statistical modified entropic gravity as a theoretical basis  for MOND}

\author{ Ehoud Pazy }

\affiliation{Department of Physics, University of connecticut, Storrs CT 06269 USA \footnote{On leave from Department of Physics NRCN, P.O.B. 9001, Beer-Sheva 84190, Israel}}
%\email{pazy@phys.uconn.edu}

\numberwithin{figure}{section}

\begin{abstract}
Considering the quantum statistics of the degrees of freedom on the holographic screen it is shown that the ratio of the number of excited bulk degrees of freedom to the number of excited surface degrees of freedom, is given by the MOND interpolating function $\tilde{\mu}$. This relationship is shown to hold also in AQUAL, and in the extension of MOND to de Sitter space. Based on the relationship between the entropy, and the number of degrees of freedom on the holographic screen, a simple expression, relating the MOND interpolating function to the ratio of the two-dimensional entropy to Bekenstein-Hawking entropy, is obtained. In terms of this expression MOND can be viewed as a modification of gravity arising due to a bound on the maximum entropy for the holographic screen.
 \end{abstract}
 
 \maketitle
 
\section{Introduction}
\label{sec:intro}

The connection between gravity and thermodynamics was first noted in the context of black hole thermodynamics in the pioneering works of Bekenstein \cite{Bekenstein} and Hawking \cite{Hawking}. It was Jacobson \cite{Jacobson} who later demonstrated that the Einstein field equations could be obtained from the first law of thermodynamics on a Rindler casual horizion, employing the connection between acceleration and temperature derived by Unruh \cite{Unruh}. Based on this connection between gravity and thermodynamics an emergent perspective of gravity has been studied by Padmanabhan (For reviews see \cite{Padmanabhan10} and \cite{Padmanabhan11}) and Verlinde \cite{Verlinde11}, as well as by others \cite{entropy}, the original idea of emergence dating back to Sakharov \cite{Sakharov}. However if one seriously wishes to consider gravity as an emergent phenomena based on the  thermodynamical formulation, the quantum statistics of the underlying degrees of freedom on the holographic screen should be considered in the low temperature limit. In Ref. \cite{Pazy12} this was done, and it had been shown that if one considers the quantum statistics of the degrees of freedom on the holographic screen, the resulting non-relatvistic dynamics is expressed by modified Newtonian dynamics  (MOND) \cite{Milgrom83}, due to Milgrom. Accordingly, in this quantum statistical formulation of gravity, there is no need to invoke dark matter to explain the observational discrepancies for galactic and extragalactic systems. Rather one needs to modify the gravitational potential for low accelerations, where the modification is due to quantum statistics. In this formulation, which shall be referred to as the quantum statistical modified entropic gravity (QSMEG), the MOND characteristic acceleration scale $a_0$, was shown \cite{Pazy12} to be related to the Fermi energy for fermionic degrees of freedom on the holographic screen and its bosonic analog for bosons \cite{remark}. The MOND interpolating  function $\tilde{\mu}$, which is a function of the ratio of the acceleration $a$ to $a_0$, was also obtained, analytically from the underlying quantum statistics. Thus through QSMEG the high acceleration limit, $a/a_0>>1$,  corresponds to the high temperature limit in which the Maxwelll Boltzmann distribution can be used, which in turn corresponds to Newtonian dynamics. Whereas in the low temperature limit one needs to consider MOND, the corrections to Newtonian dynamics arising due to quantum statistics. 

In this paper we  expand on the basic formulation of QSMEG presented in \cite{Pazy12}. We formulate QSMEG in terms of the number of bulk and surface degrees of freedom, which allows us to express the MOND interpolating function $\tilde{\mu}$, as a simple ratio between the number of bulk and surface degrees of freedom or as a ratio between the two dimensional screen entropy to Bekenstein-Hawking entropy. We also contrast  our results to those obtained by Padmanabhan \cite{Padmanabhan12} in the frame work of emergent gravity. Specifically we note the differences arising due to considering the quantum statistics of the underlying degrees of freedom on the holographic screen. Furthermore we extended QSMEG to a relativistic formulation. Sec. \ref{sec:asympt} briefly reviews the derivation of QSMEG presented in Ref. \cite{Pazy12}, and as an example we obtain through QSMEG the asymptotic MOND velocity for a rotating stellar object in the deep MOND limit (DML). In Sec. \ref{sec:QSmicro}, QSMEG is formulated in terms of the effective number of bulk and surface degrees of freedom and it is shown that one can express the MOND interpolating function as the ratio between the two. The same result is then obtained by employing  the Lagrangian formulation of MOND and then taking the DML. A calculation of the surface energy corresponding to the holographic screen in the low temperature limit is preformed, and compared  to the calculation done in Ref.\cite{Padmanabhan12} in which equipartition was assumed. Sec. \ref{sec:cosmological} describes a cosmological extension of QSMEG to de Sitter space. In  Sec. \ref{sec:Rindler}, QSMEG is considered  for a Rindler horizon and an expression for the MOND interpolating function, based on an entropy ratio, is obtained. A short discussion of the major results, and conclusions are then presented in Sec. \ref{sec:conclusions}.

\section{Obtaining the asymptotic MOND velocity through quantum statistics}
\label{sec:asympt}

To consider the underlying quantum statistics of the holographic screen degrees of freedom, one can simply modify the thermodynamical gravity formalism introduced by Verlinde \cite{Verlinde11}, by replacing the equipartition, with the relevant quantum statistical expression \cite{Pazy12}. We note in passing that the same results can be obtained alternatively, by considering the entropy of the holographic screen as was done by Padmanabhan \cite{Padmanabhan}, only this time modifying the entropy due to the underlying quantum statistics.

We now briefly describe the derivation of QSMEG as presented in  Ref. \cite{Pazy12}.  
The derivation deployed the Unruh relation between temperature and acceleration
\begin{equation}
\label{eq:Unruh}
k_BT={1 \over 2\pi}{\hbar a \over c},
\end{equation}
where $c$ is the speed of light and $k_B$ is Boltzmann's constant. For convenience we will from here on consider energy units such that $k_B=1$. The Bekenstein relation between the number of degrees of freedom on the Horizon $N$, and the area of Horizon $A$, was used. This relation, originally developed for black holes was extended to the Holographic screen of an accelerating particle 
\begin{equation}
\label{eq:Area}
N={A c^3 \over G \hbar},
\end{equation}
where $G$ is Newton's gravitational constant. 
The Einstein mass $M$, and energy $E$ relation, $E=Mc^2$ was also employed.
However the thermodynamical equipartition relation $E=(1/2)N T$ was modified by considering the quantum nature of the degrees of freedom on the holographic screen. For the low temperature limit  a Fermionic system's thermal energy is given by 
\begin{equation}
\label{eq:thermal_en}
E_{th}={T^2 N \pi^2 \over 12 E_F},
\end{equation}
where $E_F$ is the Fermi energy.  It is this change, (\ref{eq:thermal_en}), from the formalism presented in  \cite{Verlinde11} that leads to MOND rather then Newtonian gravity. One simply uses Eq. (\ref{eq:thermal_en}) transforms the temperature to acceleration through the Unruh relation, (\ref{eq:Unruh}), the energy to mass through the Einstein relation, and the number of degrees of freedom to the holographic screen area, via Eq.(\ref{eq:Area}).
Accordingly by comparing the resulting equation to MOND, the characteristic acceleration scale $a_0$ can be the identified with, 
a typical energy scale which divides between the quantum and the classical regimes, which is the Fermi energy  \cite{Pazy12}, for the fermionic case,
\begin{equation}
\label{eq:Fermi_ene}
a_0=({12 c \over \hbar \pi})E_F.
\end{equation}
Moreover one can show in the general case the MOND interpolating function is given by the ratio of the energy to its equipartition value $NT/2$. Which in turn one can analytically calculate, in terms of the dilog function $Li_2(y)$
\begin{equation}
\label{eq:mu_analytical}
{\tilde \mu}({T \over T_0})=-{6 \over \pi^2}{T  \over T_0} Li_2(-e^{{\mu}/T})- {\pi^2 \over 6}{T_0 \over T},
\end{equation}
where $ \mu$ is the chemical potential, not to be confused with $\tilde{\mu}$, the MOND interpolating function, and $T_0=\hbar a_0/2\pi c$, is the thermal equivalent of $a_0$ obtained by
the Unruh relation, Eq. (\ref{eq:Unruh}).

As an example of the connection between QSMEG and MOND we now demonstrate, how the  MOND asymptotic velocity  $v_{\infty}$ (for a recent review of MOND see \cite{MONDRev}), is obtained in the DML via the quantum statistics on the holographic screen. The DML, i.e. $a/a_0<<1$, for which $\tilde{\mu}(x) \approx x$, corresponds to low temperatures in the thermodynamical analog for gravity. The thermal energy for a two dimensional fermionic gas in the low temperature limit  is given by Eq. (\ref{eq:thermal_en}) which can be connected to accelerated motion, by converting  the temperature to acceleration through the Unruh relation, Eq. (\ref{eq:Unruh}). For the case of circular motion in an axisymmetric potential, the acceleration is given by $a=v^2/R$, where $v$ is the velocity of the stellar object and $R$ is the radius of the circular rotation. Thus for centripetal acceleration  using the Einstein energy mass equivalence relation, for the thermal energy, Eq. (\ref{eq:thermal_en}), can be written as,
 \begin{equation}
 \label{eq:thermal_energy}
Mc^2={N \pi^2 \over 12 E_F} {1 \over 4 \pi^2} {\hbar^2 v^4 \over R^2 c^2}.
\end{equation}
Expressing the number of degrees of freedom as a function on the holographic screen area $A=4 \pi R^2$, via Eq. (\ref{eq:Area}) we obtain  
\begin{equation}
\label{eq:velocity}
v^4=G M {12 c \over \pi \hbar} E_F.
\end{equation}
Identifying the Fermi energy with the MOND acceleration scale through Eq. (\ref{eq:Fermi_ene})
the expression for the MOND asymptotical velocity  is obtained
\begin{equation}
\label{eq:asy_vel}
v_{\infty}={(GM a_0)}^{1\over 4}.
\end{equation}
A further remark regarding DML in connection with QSMEG is due. It has been stressed by Milgrom the DML is defined through a scale invariance, i.e the invariance of the equations of motion under $(t,r)\rightarrow (\lambda t,\lambda r)$, in Ref. \cite{Milgrom09}  he even suggested that this invariance be the definition for the DML, instead of the definition based on the low-acceleration behavior of the MOND interpolating function, $\tilde{\mu}(x) \approx x$. On the QSMEG side, when considering the  quantum statistics of the degrees of freedom on the holographic screen the scale invariance arises quite naturally since in the low temperature limit the two dimensional grand canonical potential reduces to a power-law in the temperature. Since the grand canonical potential has no physical scale dependence all thermodynamical functions will not have one as well.

\section{ The number of effective surface and bulk degrees of freedom due to quantum statistics}
\label{sec:QSmicro}

Ref.  \cite{Padmanabhan12} considers the emergent perspective of gravity, it was stressed there, that if temperature and heat  can be related to spacetime, spacetime should  therefore have a microstructure. The  equipartition was then assumed, i.e,  relating an energy of $T/2$, to each degree of freedom on the holographic screen. Through equipartition an equality $N_{bulk}=N_{sur}$, between the number of degrees of freedom residing in the bulk $N_{bulk}$, to those residing on the surface of the holographic screen $N_{sur}$, was obtained. We shall refer to this equality as Ref.  \cite{Padmanabhan12} does, as the holographic equipartition.

As discussed in the previous section considering the quantum statistics of the degrees of freedom on the screen one has to distribute the thermal energy \cite{remark2} accordingly, which leads to the following relation obtained in \cite{Pazy12}
\begin{equation}
\label{eq:energy}
E_{th}={1 \over 2} \tilde{\mu} N T,
\end{equation}
where $N=N_{sur}$, is the number of degrees of freedom on the holographic screen. 

Starting from the above relation (\ref{eq:energy}) replacing the equipartition, we now determine how the holographic equipartition is modified.
Considering  as in \cite{Padmanabhan12}, the number of bulk states to be given by 
\begin{equation}
\label{eq:Nbulk}
N_{bulk}={E_{th} \over (T/2)},
\end{equation}
thus according to QSMEG  holographic equipartition no longer generally holds i.e, there is no equality between the number of  bulk and surface degrees of freedom. Rather,
\begin{equation}
\label{eq:degrees_of_freedom}
{N_{bulk}\over N_{sur}}=\tilde{\mu}({T \over T_0}).
\end{equation}
Eq. (\ref{eq:degrees_of_freedom}) is a major result of QSMEG, it has been obtained by considering  quantum statistics on the holographic screen. The ratio between the number of bulk degrees of freedom to the number of surface degrees of freedom  depends on the temperature. Since the ratio is given by the MOND interpolating function, then in the high temperature/acceleration limit for which  $\tilde{\mu} (x) \rightarrow 1$, the number ratio goes to unity  thus giving holographic equipartition as considered in \cite{Padmanabhan12}. 

\subsection{Connection between the number of surface and bulk degrees of freedom obtained from AQUAL}
\label{sec:equal}

In this section starting from a Lagrangian formalism for MOND, we show how Eq. (\ref{eq:degrees_of_freedom}) can be obtained. It should be noted that the  reverse inference, i.e.
obtaining the MOND Lagrangian from thermodynamic considerations can be found in Ref. \cite{Klinkhamer12}. We start by considering the action associated with MOND. In order to modify gravitation according to MOND one needs to change the gravitational action, thus modifying the gravitational Poisson equation. Such a modified action theory, was devised by Milgrom and Bekenstein \cite{Bekenstein_Milgrom}, and since the resulting Lagrangian is non-quadratic in the gradient of the gravitational potential, $ \Phi$, it was named the aquadratic lagrangian theory (AQUAL).

By varying the AQUAL action with respect to $\Phi$, one obtains, the following modified Poisson equation
\begin{equation}
\label{eq:Aqual}
\nabla \cdot  [  \tilde{\mu} (\mid \nabla \Phi \mid/a_0)\nabla \Phi  ] =4\pi G \rho,
\end{equation}
where, $\rho$ is the mass density.
Integrating over the volume and multiplying by $c^2$,
\begin{equation}
\label{eq:bulk_energy}
Mc^2={c^2 \over 4 \pi G} \int dV \nabla \cdot \left[{\tilde{\mu}} (\mid \nabla \Phi \mid/a_0)\nabla \Phi \right ],
\end{equation}
the energy $E_{th}$ is obtained, through the Einstein mass energy relation.
Employing Gauss's theorem to the above Eq. (\ref{eq:bulk_energy}),
\begin{equation}
\label{eq:Gauss}
E_{th}={c^2 \over 4 \pi G} \int_{\partial V} dA (-\hat{n} \cdot \nabla \Phi ) \tilde{\mu} (\mid \nabla \Phi \mid/a_0).
\end{equation}
Relating the gradient of the potential to the acceleration and using the Unruh temperature relation, Eq. (\ref{eq:Unruh}), the following expression for the thermal energy is derived
\begin{equation}
\label{eq:energy_temp}
E_{th}= \int_{\partial V} {dA \over (G\hbar/c^3)} \left [ {1\over 2} T \tilde{\mu} (\mid T/T_0 \mid) \right].
\end{equation}
From the above expression, Eq. (\ref{eq:energy_temp}), integrating over a sphere at a constant temperature and employing
\begin{equation}
\label{eq:sphere}
N_{sur}=\int_{\partial V} {dA \over (G\hbar/c^3)},
\end{equation}
we obtain again, that the ratio between the number of degrees in the bulk to the number of degrees on the surface of the sphere, is equal to the MOND interpolating function i.e, Eq. (\ref{eq:degrees_of_freedom}).

So far the thermal energy was associated with thermally exciting degrees of freedom on the holographic screen. In the next section we consider the non-thermal energy associated with the screen, which is typically associated with dark matter in the $\Lambda$CDM model of cosmology.

\subsection{ The DML surface energy  } 
\label{sec:NDML}
 
In this section we calculate the energy of the  surface degrees of freedom in the low temperature limit. To do so we start off with the DML dynamics  for a rotating stellar object 
\begin{equation}
\label{eq:DML}
{\left ({v^2 \over R} \right )}^2={G M \over R^2}a_0,
\end{equation}
and calculate the gravitational energy \cite{Verlinde_priv}. We introduce  $\mid \nabla \Phi_M \mid^2={G M a_0 /R^2}$, where the index $M$ is used to indicate that this gravitational potential originates from MOND. We  calculate the gravitational energy associated with the gravitational field
\begin{equation}
\label{grav_ene}
E_G(R)={1\over 8 \pi G} \int dV \mid \nabla \Phi_M \mid ^2,
\end{equation}
obtaining
\begin{equation}
\label{eq:Grav_ground_ene}
E_G(R)= {M R c \over \hbar} {\hbar \over 2 c} a_0,
\end{equation}
where the ratio, $\hbar /c$, was intorduced in order to convert $a_0$ to a temperature scale through the Unruh relation (\ref{eq:Unruh}). It should be noted that according to calculations made by Verlinde \cite{Verlinde_priv} one can determine through Eq. (\ref{eq:Grav_ground_ene}), the ratio of dark matter $M_D$ component, percentage in the Universe, to  Baryonic matter $M$ percentage, given the percentage of Baryonic matter in the Universe. His calculation, agreeing with observational data, estimates the dark matter component to be $22.5 \% $ for  $4\%$ baryonic matter. This estimate is achieved by equating the left hand side of Eq. (\ref{eq:Grav_ground_ene}), with the integral over the volume of $G M_D^2/R^2$. 

We take a different approach, transforming $a_0$ in Eq. (\ref{eq:Grav_ground_ene}) to a temperature scale,  
\begin{equation}
\label{eq:Grav_Temp}
E_G(R)= \pi {M R c \over \hbar}  T_0.
\end{equation}
We obtain a gravitational energy, Eq. (\ref {eq:Grav_Temp}),  which does not depend on the temperature $T$, rather it depends on some fixed temperature scale $T_0$. If this temperature scale is related to a  typical energy scale, which divides between the quantum and the classical regimes, the Fermi energy $E_F$, for  fermionic degrees of freedom on the holographic screen, as was suggested in Ref \cite{Pazy12}, then Eq. (\ref{eq:Grav_Temp}) is the non-thermal energy of the holographic screen which is quantum mechanical in its nature. The above result  can also be viewed according to the ideas presented in Refs. \cite{Pikhitsa, Klinkhamer11} in which the MOND acceleration scale $a_0$ is considered to be a minimal temperature for the microscopic degrees of freedom on the holographic screen. In this view the energy related with the screen is due to a coupling with a cosmological heat bath with a temperature, $T_0$.

To further understand the meaning of Eq. (\ref{eq:Grav_Temp}), we compare it to a similar calculation in which holographic equipartition between the number of bulk and surface degrees of freedom was assumed. Through the comparison it will be shown that essentially Eq. (\ref{eq:Grav_Temp}) states that non-thermal, ground state,  energy of the system is the equilvelent of each surface degree of freedom having an average energy of $T_0/2$. Again this result can be viewed in the two ways described above.

In \cite{Padmanabhan12},  the kinetic energy $E_{kin}=(1/2)Mv^2$  was calculated, where $v={(GM/R)}^{1/2}$, after which the virial theorem was employed to obtain the gravitational energy in terms of the Unruh temperature, as function of the radius in the non-relativstic limit
 \begin{equation}
\label{eq:Pad_grav_ene}
E_G(R)=2 \pi {M R c \over \hbar} T.
\end{equation}

Eq. (\ref{eq:Pad_grav_ene}) has a rather intuitive meaning when, using the definition \cite{Padmanabhan12}
\begin{equation} 
\label{eq:Neff}
N_{eff}\equiv{v^2 \over 2 c^2} N_{sur} =2 \pi (MRc/ \hbar),
\end{equation}
where $N_{sur}=4\pi R^2 /L_p^2$ and $L_p^2=Gh/c^3$. It simply states that
\begin{equation}
\label{eq:gs}
E_G=N_{eff} T,
\end{equation}
i.e. each effective degree of freedom has an average energy of $T$.  Eq. (\ref{eq:Pad_grav_ene}) was calculated on basis of holographic equipartition, i.e  $N_{bulk}=N_{sur}$, thus $N_{eff}$ in Eq. (\ref{eq:gs}), can be considered as $N_{bulk}$. If however we wish to calculate the energy associated with the surface degrees of freedom we need according to Eq. (\ref{eq:degrees_of_freedom})  to divide the energy by $\tilde{\mu}$ . Dividing $N_{eff}$ in Eq.(\ref {eq:gs}) by $\tilde{\mu}$ one obtains in the DML Eq. (\ref{eq:Grav_Temp}). Thus Eq. (\ref{eq:Grav_Temp}) can be viewed as an expression for the surface energy in the low temperature limit, in which each degree of freedom has effectively an average energy of $T_0/2$.

The non-thermal energy Eq. (\ref{eq:Grav_Temp}), does not play a role in the classical limit for which only the thermal energy $E_{th}$ should be considered, however it is vital for gravitational lensing and once cosmological scales are considered one can not ignore it.

\section{QSMEG in  the cosmological case}
\label{sec:cosmological}

Previous sections have considered the modifications of the classical gravity regime due to QSMEG. It is however well known that the MOND characteristic acceleration scale $a_0$ can be simply related to cosmological scales, $a_0\approx c  H_0 /2 \pi\approx 2(\Lambda/3)^{1/2}$, where $\Lambda$ is the cosmological constant. This hints at the possibility of $a_0$ having cosmological significance. In the following sections we will show how our results, specifically Eq. (\ref{eq:degrees_of_freedom}), extend to the relativistic regime and attempt to connect them with cosmological extensions of MOND. Though QSMEG requires only an horizon and thus can be easily extended to cosmology, the same can not be said for MOND \cite{Starkman} which was originally devised as a phenomenological theory introduced to solve discrepancies on the galactic scale.
There have been quite a few attempts to construct a cosmological model for MOND, most of which are based on introducing extra fields \cite{MONDRev}, due to this reason we will not consider them here.  However there have also been some attempts to construct a cosmological extension of MOND based on the thermodynamical description of gravity. In Ref. \cite{Neto}, $a_0$ was associated with a critical temperature, and a MOND like Friedmann equation was obtained, however this equation depends on the radial co-moving coordinate. In Ref. \cite{Zhang} a  MOND like cosmology was developed based on a MOND Friedmann like equation and in \cite{Kiselev} an homogenous extrapolation of MOND to cosmology was defined by considering a cosmological scaling for $a_0$, instead of it being a constant. There have also been a few suggestions to extends MOND to the cosmological regime by associating  $a_0$ with the Unruh temperature of the de Sitter horizon Refs. \cite{Pikhitsa,Klinkhamer11,MOND_de_Sitter}, as originally proposed by Milgrom \cite{Milgrom99} for the de Sitter space. In the following section we will use this de Sitter MOND version and demonstrate it is consistent  with our previous results, specifically Eq. (\ref{eq:degrees_of_freedom}).
 
\subsection{ Number of surface and bulk degrees of freedom for de Sitter space}
\label{sec:sub_cosmo}
In this section we start by following Ref. \cite{Padmanabhan12}, we briefly review the thermodynamical description of the physics of de Sitter space, in which a temperature is associated to the horizon and one assumes holographic equipartition, $N_{bulk}=N_{sur}$. We will then modify the results according to the MOND de Sitter version \cite{Milgrom99}, and thus verify the consistency of QSMEG applied to de Sitter space. 

In \cite{Padmanabhan12} a de Sitter Universe with a Hubble constant $H$ was considered, and the  holographic equipartition condition, $N_{bulk}=N_{sur}$, was shown to reduce to 
\begin{equation}
\label{eq:deSitter}
H^2={8\pi L_p^2 \rho  \over 3},
\end{equation}
where $\rho$ is the energy density.

For a de Sitter Universe, the number of degrees of freedom attributed to a spherical surface with a Hubble radius $H^{-1}$, is given by
\begin{equation}
\label{eq:Nsurfde}
N_{sur}={4\pi \over (L_p^2 H^2)},
\end{equation}
and the number of bulk degrees of freedom $N_{bulk}$ is calculated by dividing the Komar energy $\mid (\rho+3P)\mid V$, by $(1/2) T$,  where $P$ is  the radial pressure. For a pure de Sitter space, for which $P=-\rho$ ,
\begin{equation}
\label{eq:Nbulkde}
N_{bulk}={4 \rho V \over T},
\end{equation}
where the volume is the proper volume of a Hubble sphere $V=4\pi/3 H^3$, and the temperature is the Horizion temperature 
\begin{equation}
\label{eq:Hor_temp}
T=( H/2 \pi).
\end{equation}

We now consider a MOND formulation for de Sitter space which was originally proposed by Milgrom \cite{Milgrom99} and later developed by others in the context of Verlinde's thermodynamical approach \cite{Pikhitsa,Klinkhamer11,MOND_de_Sitter}. In this formulation the Unruh temperature difference  $\Delta T$ measured by a non-interial observer with an acceleration $a$, corresponding to an Unruh temperature $T$,  is measured with respect to a background temperature $T_\Lambda=(\Lambda/3)^{1/3}/2\pi$. This background temperature is measured by all intertial observers. With respect to the background temperature, the temperature difference is
\begin{equation}
\label{eq:MOND_de_Sitter}
\Delta T= T\tilde{ \mu} (T/T_\Lambda).
\end{equation}
To obtain the total thermal energy one should multiply the above Eq. (\ref{eq:MOND_de_Sitter}) by $N/2$ \cite{Klinkhamer11}, obtaining Eq. (\ref{eq:energy}). Since Eq. (\ref{eq:energy}) leads to Eq. (\ref{eq:degrees_of_freedom}), we now modify the number of bulk states accordingly to the available energy according to  Eq. (\ref{eq:degrees_of_freedom}), then by dividing, Eq.(\ref{eq:Nbulkde}) by Eq.(\ref{eq:Nsurfde}), we obtain instead of the standard result, Eq. (\ref{eq:deSitter}),
\begin{equation}
\label{eq:ModdeSitter}
H^2={8 \pi L_p^2 \rho \over 3 \tilde{\mu} (T/T_\Lambda)}.
\end{equation}
It should be noted that in order for Eq. (\ref{eq:ModdeSitter}) to be independent of the radial co-moving coordinate as should be expected in a homogenous Universe, one can consider that instead of $a_0$ being a constant, it scales with the expansion factor in the same way as the acceleration $a$. This idea  was considered in \cite{Kiselev}, and was found to fit well with observational data, without any need for dark matter.

Eq. (\ref{eq:ModdeSitter}) can also be expressed as a relation between the energy and the number of degrees of freedom on the Horizon. By integrating the density over the volume we obtain the Komar energy for a de Sitter Universe  
\begin{equation}
\label{eq:ene_de_Sitter}
E=\int dV {1\over 4 \pi L_P^2} 3  \tilde{\mu} (T/T_\Lambda) H^2.
\end{equation}
Considering the spherical symmetric case the we obtain
\begin{equation}
\label{eq:surf_ene}
E={1 \over 4\pi}{A \over L_p^2}   \tilde{\mu} (T/T_\Lambda) H,
\end{equation}
where $A=4\pi/H^2$. Expressing the Hubble constant in terms of the  temperature (\ref{eq:Hor_temp}), we obtain
\begin{equation}
\label{eq:ene_de_Sitter_temp}
E={1 \over 2}{A \over L_p^2}   \tilde{\mu} (T/T_\Lambda) T.
\end{equation}

Noting that  $A/L_p^2=N_{sur}$, Eq. (\ref{eq:ene_de_Sitter_temp}) is equivalent to Eq. (\ref{eq:energy}) demonstrating that QSMEG can be successfully applied to de Sitter space. However Eq. (\ref{eq:ene_de_Sitter_temp}) is not specific to de Sitter space, rather it is a result of considering the quantum statistics on a horizon, as such it should also apply to Rindler horizons as well.

\section{Number of surface and bulk degrees of freedom for a Rindler horizon}
\label{sec:Rindler}
In this section we will determine the effects of considering the quantum statistics of the degrees of freedom on a Rindler horizon. A task which in the general case requires a relativistic formulation. Unfortunately we are unable to compare the results to a  covariant MOND theory, since current covariant theories which reduce to MOND in the quasi-static, weak-field limit, include new fields \cite{MONDRev}. To  proceed we consider a modification of the method employed by Padmanabhan \cite{Padmanabhan04,Padmanabhan10b}  to obtain holographic equipartition. We modify his results in accordance with QSMEG, specifically according to Eq. (\ref{eq:degrees_of_freedom}), and obtain a simple expression for the MOND interpolating function in terms of entropy.

\subsection{The holographic equipartition from Einstein's field equations}
\label{sec:equi}
We start by briefly reviewing the calculations and results  of \cite{Padmanabhan04,Padmanabhan10b} in which the entropy for a static horizon was defined and compared to the energy obtaining equipartition. The entropy definition was given for a static  metric  $ds^2=-N^2dt^2+\gamma_{\mu \nu}dx^{\mu}dx^{\nu}$, where $N$ and $\gamma_{\mu \nu}$ are independent of the time, $t$. The $-+++$ signature is used, as well as units for which $c=\hbar=k_B=1$; the Greek indices corrospond to indices $1,2, 3$ . The comoving observers at $x^{\mu}=constant$ have the four acceleration $a^i=(0,\partial^{\mu}N/N)$. An horizon is obtained once $N\rightarrow 0$, on a two surface defined by, $N^2a^2\equiv(\gamma_{\mu \nu}\partial^{\mu}N\partial^{\nu}N)$. Using a coordinate transformation, from the original coordinates $x^\mu$, to the set $(N,y^A),A=2,3$, where the $y^A$ coordinates denote the two transverse coordinates to the $N=constant$ surface. The metric, in these coordinates, can be written as
\begin{equation}
\label{eq:metric}
ds^2=-N^2dt^2+{dN^2 \over(Na)^2}+\sigma_{AB}\left (dy^A-{a^A dN \over Na^2} \right) \left (dy^B-{a^B dN \over Na^2} \right).
\end{equation}

The entropy is then defined as
\begin{equation}
\label{eq:h_ent}
S={1 \over 8 \pi G} \int \sqrt{-g} d^4 x \nabla_ia^i,
\end{equation}
where $g=det(g_{\mu \nu})$, of the metric defined in Eq. (\ref{eq:metric}), and after performing the integration over $t$ and the radial coordinate it is given by
\begin{equation}
\label{eq:horizon_ent}
S={\beta \over 8 \pi G} \int_{\partial V} \sqrt{\sigma} d^2 x(Nn_{\mu}a^{\mu}),
\end{equation}
where the integration is over $\partial \nu$ which is the surface of some compact volume $\nu$, and $n_\mu$ is the normal to $\partial \nu$,  $\sigma=det(\sigma_{AB})$ and $\beta=2 \pi/ \kappa$ is the horizon temperature defined through the surface gravity $\kappa$. Defining the energy in  the region $\nu$,  as
\begin{equation}
\label{eq:energy_space_time}
E= 2\int_{\nu} d^3x \sqrt{\gamma} N\left (T_{ab}-{1 \over 2}Tg_{ab} \right)u^a u^b,
\end{equation}
with $T_{ab}$  the energy momentum tensor, one obtains equipartition, Ref. \cite{Padmanabhan10b}. This is achieved by first using Einstein's field equations, through which the divergence of the acceleration is related to the source
\begin{equation}
\label{eq:Einstein}
D_\mu (N a^\mu)\equiv 8 \pi N\left (T_{ab}-{1 \over 2}Tg_{ab} \right)u^a u^b,
\end{equation}
where $D_\mu$ is a covariant derivative operator corresponding to the 3-space metric. Then employing Gauss's theorem
\begin{equation}
\label{eq:energy_relativ}
E={1 \over 2} \int_{\partial \nu} {\sqrt{\sigma} d^2x \over L_p^2}{N a^\mu n_\mu \over 2 \pi}.
\end{equation}
Defining an effective local temperature $T_{loc}=(N a^\mu n_\mu /2 \pi)$. Equipartition is thus obtained
\begin{equation}
\label{eq:equipartition_rel}
E={1 \over 2} \int_{\partial \nu} dn T_{loc},
\end{equation}
where $dn=\sqrt{\sigma}d^2x/ L_p^2$. Furthermore by comparing the energy Eq. (\ref{eq:energy_relativ}) with the entropy Eq. (\ref{eq:horizon_ent}), on the horizon, the following thermodynamic relation
\begin{equation}
\label{eq:thermodynamic}
E=2ST,
\end{equation}
is obtained (for an alternative derivation see \cite{Banerjee}). 

\subsection{Modifying the equipartition and entropic considerations}
\label{sec:mod_equip}
 In this section we modify the relations obtained in the previous section according to QSMEG.
 First, we naively modify the equipartition obtained in Eq. (\ref{eq:equipartition_rel}),
 according to Eq. (\ref{eq:energy}) thus obtaining 
\begin{equation}
\label{eq:MOND_rel}
E={1 \over 2} \int_{\partial \nu} \tilde{\mu}  dn T_{loc}.
\end{equation}
 We then use Eq. (\ref{eq:thermodynamic}), to relate the entropy to the MOND interpolating function,
\begin{equation}
\label{eq;entropy_mod}
S={1 \over 4}\int_{\partial \nu} \tilde{\mu}  dn.
\end{equation}

Based on Eq. (\ref{eq;entropy_mod}) a different expression defining the MOND interpolating function can be given, instead of Eq.(\ref{eq:energy}), which defines $\tilde{\mu}$ through the ratio of the number of degrees on the holographic screen to the number of degrees of freedom in the bulk, one can use an entropy based definition. Given that the number of degrees of freedom is four times the Bekenstein-Hawking entropy \cite{Bekenstein, Hawking}
\begin{equation}
\label{eq:entropy_BH}
S_{BH}={Ac^3 \over4 G \hbar},
\end{equation}
it is possible through Eq. (\ref{eq;entropy_mod}) to express the MOND interpolating function as
\begin{equation}
\label{eq:mu_ent_ratio}
\tilde{\mu} ({T \over T_0})={S \over S_{BH}}.
\end{equation}
Apart from it's simple form, Eq. (\ref{eq:mu_ent_ratio}) connects to the thermodynamic interpretation of gravity as a entropic force, as was put forward by Verlinde \cite{Verlinde11}.  Employing this view the gravitational force being entropic is modified since there is a bound on the maximum entropy of the holographic screen.

In the light of these entropic considerations one can restate the low temperature limit (or non-thermal) for the surface energy Eq.(\ref{eq:Grav_Temp}), noting that according to the Bekenstein entropy bound \cite{Bekenstein81}
\begin{equation}
\label{eq:bound}
S\leq {2 \pi R M c \over \hbar}.
\end{equation}
Eq. (\ref{eq:Grav_Temp}) can be rewritten as
\begin{equation}
\label{eq:Ene_bound}
E_G= {S_{max} T_0 \over 2},
\end{equation}
where $S_{max}$ is the maximal entropy corresponding to an equality in the Bekenstein bound, Eq. (\ref{eq:bound}). In this context, it is clear that $E_G$ represents the heat content of the holographic screen \cite{Padmanabhan13}. The above result (\ref{eq:Ene_bound}) can be understood in terms of the ideas presented by \cite{Pikhitsa,Klinkhamer11} in which the MOND acceleration scale $a_0$ is interpreted as corresponding to a minimal temperature $T_0$. The entropy then reaches its maximum value at the minimal temperature available to the system. However as was previously stated, Eq. (\ref{eq:Grav_Temp}) can also be viewed as
\begin{equation}
E_G={N T_0 \over 2}.
\end{equation}
In this view the energy is quantum mechanical and at the zero temperature limit, each degree of freedom has on average half the Fermi energy (in the fermionic case).
\section{Discussion and Conclusion}
\label{sec:conclusions}

In Ref. \cite{Pazy12}  QSMEG was first considered, and shown to lead to MOND.  Here the ideas presented in Ref. \cite{Pazy12} where extended. Based on quantum statistical 
modification of  equipartition, QSMEG was shown to be consistent both with MOND and with AQUAL, as well as for the cosmological extension of MOND to de Sitter space. 

Eqs. (\ref{eq:degrees_of_freedom} ,\ref{eq:mu_ent_ratio}),  which give simple expressions for the MOND interpolating function $\tilde{\mu}$, are the main results of this paper. In Ref. \cite{Pazy12} MOND was introduced as a result of considering the quantum statistics of the degrees of freedom on the holographic screen. Based on Eq. (\ref{eq:mu_ent_ratio}), MOND can now also be viewed as a modification of gravity, being an entropic force, due to a bound on the maximum entropy of the horizon degrees of freedom. Thus MOND which is defined in terms of an acceleration scale $a_0$, and an interpolating function $\tilde{\mu}$, can now be related in a thermodynamic formulation to a minimal temperature or energy scale (Fermi energy for the fermionic case) corresponding to $a_0$ \cite{Pikhitsa,Klinkhamer11} and a maximal entropy for the holographic screen, related to $\tilde{\mu}$. Based on these relations the expression, Eq. (\ref{eq:Ene_bound}), for the surface energy, gets new meaning. It can be understood as the heat content of the holographic screen which is given in terms of the maximum entropy times the minimal temperature.

For the classical non-reltavistic dynamics there is no difference between  fermionic and bosonic degrees of freedom on the holographic screen, due to the remarkable thermodynamical equivalence of Fermi and Bose gases in two dimensions. It can be shown that the two gases have the same specific heat and even the same entropy \cite{Lee02}, however the two gases differ in their total energy since the Fermi gas ground state energy is $E_{G}=(1/2)N E_F$. Since in the relativistic case the total energy content has physical significance, the symmetry between the two gases is broken. Moreover the calculation of the surface energy of the screen for the lowest temperature possible, Eq. (\ref{eq:Grav_Temp}),  has physical significance. It was even shown that through it one can deduce the percentage of missing energy which is typically associated with "dark matter". Essentially in the QSMEG formulation this energy is the vacuum energy, which is responsible for gravitational lensing, thus reconciling MOND with gravitational lensing observations without the need for dark matter. Therefore it seems reasonable due to observations that one should consider, fermionic degrees of freedom, rather then bosonic, on the holographic screen.

In considering QSMEG in de Sitter space, a modified Friedmann equation was obtained, Eq. (\ref{eq:ModdeSitter}). This seems to indicates that if QSMEG and MOND correspond also in the relativistic limit, a covariant MOND theory should not necessarily correspond to General Relativity rather one may consider an alternative gravitational theory as a natural covariant extension of MOND.


\begin{thebibliography}{99}
\bibitem{remark}
For the two dimensional hologrphic screen the energy scale for fermions and bosons is the same since the specific heat of a two dimensional gas is the same for fermionic and bosonic degrees of freedom.
\bibitem{remark2}
We use the notation $E_{th}$ for the part of the energy which diminishes when the temperature/acceleration goes to zero.
\end{thebibliography}
\end{document}